\documentclass[twocolumn]{aa}
\usepackage{graphicx}
\usepackage{txfonts}
\begin{document}
   \title{Discovery of very nearby ultracool dwarfs from DENIS}
   \author{T.R. Kendall
          \inst{1}
          \and
          X. Delfosse\inst{1}
          \and
          E.L. Mart\'{\i}n\inst{2}
          \and
          T. Forveille\inst{3}
          } 

   \offprints{T.R.Kendall}

   \institute{Laboratoire d'Astrophysique de Grenoble, 
              Universit\'{e} Joseph Fourier,
              38041 Grenoble Cedex 9, France\\
              \email{tkendall@obs.ujf-grenoble.fr,delfosse@obs.ujf-grenoble.fr}
         \and
             Instituto de Astrof\'{\i}sica de Canarias,
             Via L\'{a}ctea, E-38200 La Laguna, Tenerife, Spain \\
             \email{ege@ll.iac.es}
          \and
              CFHT Corporation, 68--1238 Mamalahoa Highway, Kamuela, HI 96743, USA \\
             }

   \date{Received <date> / Accepted <date>}

   \abstract{We report new spectroscopic results, obtained with UKIRT/CGS4, of a 
sample of 14 candidate ultracool dwarfs selected from the DENIS (Deep Near-Infrared
Survey of the Southern Sky) database. A further object, selected from the 2MASS Second
Incremental Release, was observed at a later epoch with the same instrument.
Six objects are already known in the literature; we re-derive their
properties. A further four prove to be very nearby ($\la$\,10\,pc)
mid-to-late L-dwarfs, three unknown hitherto, two of which are almost certainly substellar. These findings increase
the number of L-dwarfs known within $\sim$\,10\,pc by $\sim$\,25\%. The remainder of the objects discussed here
are early L or very late M-type dwarfs lying between $\sim$45 
and 15\,pc
and are also new to the literature.  Spectral types have been derived by direct comparison
with {\it J-,H-} and {\it K-} band spectra of known template ultracool dwarfs given by
Leggett et al.\thanks{\tt ftp://ftp.jach.hawaii.edu/pub/ukirt/skl/dL.spectra/} For the known objects, we generally 
find agreement to 
within $\sim$1 subclass with previously derived spectral types. Distances are determined from the most recent 
M$_{\rm J}$ vs. spectral type calibrations,
and together with our derived proper motions yield kinematics for most targets consistent with that expected for
the disk population; for three probable late M-dwarfs, membership of a dynamically older population is postulated. 
The very nearby L-type objects discussed here are 
of great interest for future studies of binarity and parallaxes. 
   
   \keywords{stars: low mass, brown dwarfs -- stars: late-type -- stars: kinematics -- stars: distances --
                infrared: stars -- surveys
               }
   }

   \maketitle
%

\section{Introduction}

The analysis of current near-infrared sky surveys such as 2MASS (Two Micron All Sky Survey; \cite{skr97}),
DENIS (\cite{epc97}) and the Sloan Digital Sky Survey (SDSS; \cite{yor00}) is rapidly revolutionising
our knowledge of the very low-mass dwarf population in the solar neighbourhood. Observations have required
the establishment of new spectral classes (L,T) to characterise the very coolest dwarfs (\cite{kir99};
\cite{mar99}) and very recently \cite{cru03} have increased the number of known
M7 -- L6 dwarfs by a further 127\% (186 new objects), by exploitation of the 2MASS Second Incremental Release.
It is the clear goal of current efforts, using existing data, to produce a complete, volume-limited sample of 
ultracool dwarfs, over the whole sky. 

Such dwarfs (of spectral types M7 and later) are likely to have ages of a few Gyr and, with reference to 
theoretical models (e.g. \cite{bar03} 
and references therein) 
are likely to be substellar. Indeed, as pointed out by \cite{leg01} any object later 
than L5 has to be substellar; i.e. incapable of sustaining core hydrogen fusion at 
any point in its lifetime. The field population of L- and T-dwarfs thus represents an important link to
even less massive, younger objects known in nearby star-forming regions. 

The DENIS survey has demonstrated its ability to detect very cool stellar objects with the detection
of the first L-dwarf populations (\cite{del97}). To date, 5700 deg$^2$ of survey data have been 
explored, yielding a sample of 300 ultracool dwarf candidates selected to have $(I-J)$\,$>$\,3.0, 
complete to 
$I\sim$\,18 and reaching $I$\,=\,19.0. These objects are plotted in Fig.\,1~(crosses). 
Optical spectroscopy of the complete sample is 
underway and will be published in a future paper. In this Letter, we discuss near-infrared spectroscopy
obtained for selected relatively bright ($I\sim$\,15--17.5) objects with 3.0\,$<(I-J)<$\,4.0. The previously 
published DENIS L-dwarfs (\cite{del99}; \cite{mar99}) have $(I-J)>$\,3.1.

\begin{table*}
\begin{center}
\caption[] {Basic and derived data for observed targets. The abbreviated name Dnnnn
will be used throughout this paper. Co-ordinates are Equinox J2000 and are given in unabbreviated form. 
Previously known objects are referenced in the final column. 
$IJK$ magnitudes are from the DENIS
database and have typical errors 0.05--0.1\,mag. The modified Julian date of the DENIS observation,
galactic latitudes, derived spectral types and distances 
are given in cols. 6, 7, 8 and 9.
Spectral types quoted are the mean of those derived from independent $K$- and $H$-band derivations,
where both are available (see Col. 10).
For 2MJ1112, the $I$-band magnitude is from UKST/SuperCosmos$^1$; $JK$ from 2MASS.}  
\begin{tabular}{lllllllllll}\hline\hline
Name & DENIS-P & $I$ & $J$ & $K$ & MJD & $b$ & Sp. & d/pc & Band & ref.\\
\hline
D1048 & J104842.81+011158.2 &  16.2 & 12.9 & 11.5 & 51828.0 & +50.79 & L4   & 9.1$^{-0.9}_{+1.0}$ & $K$  & H02 \\  
D1411 & J141121.30--211950.6 &  15.5 & 12.5 & 11.3 & 51366.5 & +37.83 & M9   & 16.0$^{-0.9}_{+1.1}$ & $H,K$ & C03 \\
D1425 & J142527.97--365023.4 &  17.7 & 13.7 & 11.7 & 51828.0 & +22.32 & L5   & 10.6$^{-1.1}_{+1.2}$ & $K$ & -\\
D1456 & J145601.39--274736.4 &  16.4 & 13.2 & 12.2 & 51374.5 & +27.89 & M9   & 22.0$^{-1.3}_{+1.6}$ & $ H,K$ & C03 \\
D1510 & J151047.85--281817.4 &  16.0 & 12.8 & 11.4 & 51828.0 & +25.27 & M8   & 21.0$^{-1.5}_{+2.0}$ & $H,K$ & G02 \\
D1514 & J151450.16--225435.3 &  17.1 & 14.0 & 12.9 & 51828.0 & +29.14 & M7   & 44.1$^{-4.4}_{+6.0}$ & $K$ & - \\ 
D1539 & J153941.96--052042.4 &  17.5 & 13.8 & 12.4 & 51828.0 & +37.98 & L2   & 19.5$^{-1.5}_{+1.5}$ & $H$ & - \\
D1705 & J170548.38--051645.7 &  16.6 & 13.2 & 12.1 & 51698.6 & +20.62 & L4   & 10.7$^{-1.0}_{+1.1}$ & $H,K$ & - \\   
D2036 & J203608.64--130638.3 &  18.2 & 14.7 & 13.5 & 51828.0 &  --29.07 & M9.5 & 41.7$^{-2.4}_{+2.7}$ & $H,K$ & - \\
D2057 & J205754.10--025229.9 &  16.6 & 13.2 & 11.6 & 51786.7 & --29.26 & L1.5 & 16.2$^{-1.1}_{+1.2}$ & $H,K$ & C03 \\
D2200 & J220002.05--303832.9 &  16.7 & 13.4 & 12.4 & 51776.7 & --52.54 & L0   & 21.7$^{-1.3}_{+1.4}$ & $H,K$ & - \\ 
D2229 & J222958.15--065043.2 &  18.0 & 14.5 & 13.2 & 51828.0 & --50.80 & M9.5 & 38.0$^{-2.2}_{+2.5}$ & $H,K$ & - \\
D2252 & J225210.73--173013.4 &  17.9 & 14.2 & 12.8 & 51435.6 & --60.88 & L7.5 & 8.3$^{-0.5}_{+0.7}$  & $H,K$ & - \\
D2254 & J225451.90--284025.4 &  17.4 & 14.1 & 12.8 & 51775.8 & --64.26 & L0.5 & 28.2$^{-1.7}_{+1.8}$ & $H$ & C03 \\
\hline
2MJ1112 & 2MASS J11124910-2044315 & 18.3 & 14.9 & 13.5 & 50930.6 & +36.51  & L0.5: & 41.1$^{-10.2}_{+11.7}$ &  $J$  & -  \\
\hline
\end{tabular}
\end{center}
\begin{flushleft}
1. {\tt http://www-wfau.roe.ac.uk/sss}; C03: \cite{cru03}; G02: \cite{giz02}; H02: \cite{haw02}\\
\end{flushleft}
\end{table*}

\begin{figure}
   \centering
\includegraphics[angle=-90,width=9cm]{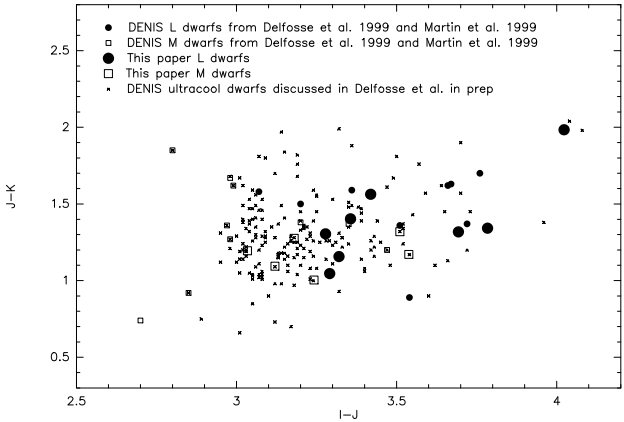}
      \caption{DENIS colour-colour diagram. The objects discussed in this Letter are represented by large
               circles (L-dwarfs) and large squares (M-dwarfs). Small symbols are the previously published
               DENIS ultracool dwarfs (\cite{del99}; \cite{mar99}). The complete sample of DENIS
               candidate ultracool dwarfs are plotted as crosses. 
                             Note that the diagram excludes objects without a DENIS-$K$ magnitude: some
                such objects are retained in our overall sample on the basis of their $(I-J)$ colour only.}
   \end{figure}

\section{Observations and Data reduction}

Near-infrared spectra were obtained on Jun 25--27, 2001, using CGS4 (Cooled Grating
Spectrometer 4) on the United Kingdom Infrared Telescope (UKIRT). Conditions
were good and the seeing was $\sim$1.1\arcsec.
The 40 mm$^{-1}$ grating was employed using the long camera yielding complete  wavelength
coverage of in the $H$ and $K$ passbands at a resolution $R$\,$\sim$\,400.
The 2MASS object, 2MJ1112, was observed in the $J$-band on 3 Feb 2003 with a similar
instrumental setup.

Data reduction was performed using the ORAC pipeline developed at UKIRT. However, telluric correction,
performed using standard (A- and F-type) spectra taken before
and after each science exposure, yielded spurious residual features 
resulting from hydrogen line absorptions in the standard spectra. Such features were fitted and removed and, after 
division by an appropriate Planck function, the divisor spectra were re-divided into the
target spectra. 
In cases where divisor spectra
suffered from fringing, a different standard spectrum (selected to be as close as possible a match 
 in observation time and airmass) was employed. This
does not affect the relative flux calibration (i.e. spectral shape) important for
spectral typing.

 \begin{figure}
   \centering
\resizebox{\hsize}{13cm}{\includegraphics[bb=0 120 543 790]{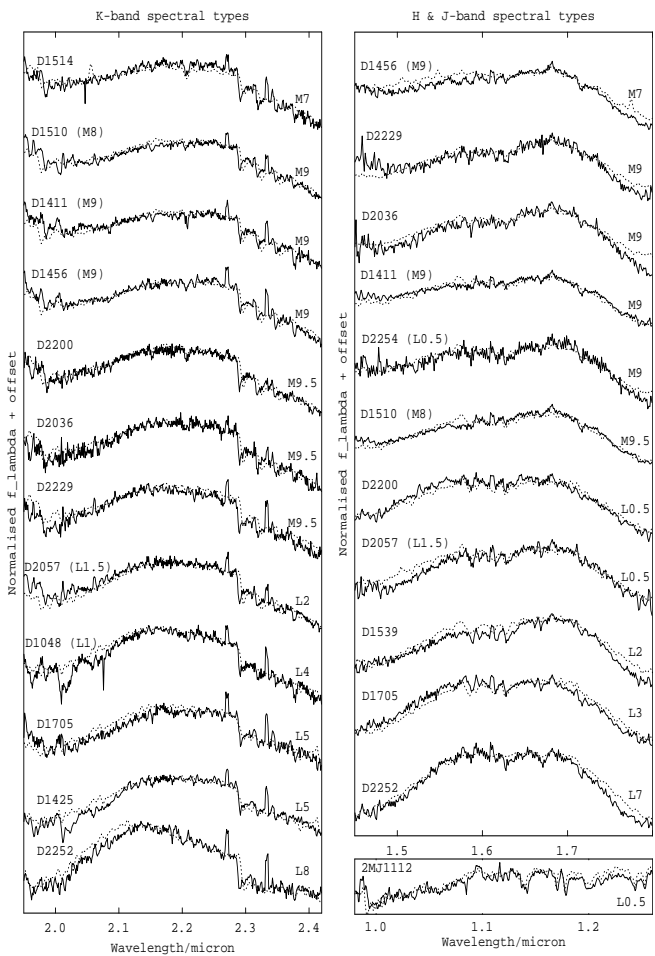}}
      \caption{Spectral typing results at R\,$\sim$\,400. Derived types are given on the right hand side of
               each panel. Previously known types are given in parentheses. In all cases, $K$-band spectra 
               have been normalised
               to the mean flux at $\sim$\,2.29\,$\mu$m, and $H$-band spectra to the highest point of the spectrum. 
               Template spectral types are on the Kirkpatrick
               et al. (1999) system (see Sect. 3). For clarity, each spectrum is offset from its neighbours
               by 0.5 continuum units.}
         \label{f1}
   \end{figure}

\section{Spectral types, distances and kinematics}

\begin{table}
\begin{center}
\caption[] {The nearest L-dwarfs to the Sun, ordered by distance. Where distances are measured from a 
parallax determination,
this is indicated. Otherwise, distances are based on current calibrations and have been obtained from the literature. 
Rather few L-dwarfs have been found within $\sim$\,10\,pc and this work increases the known database of such objects
by $\sim$\,25\%.}
\begin{tabular}{llllll}\hline\hline
Name & Survey id.\& ref. & Sp. & d/pc & Ref. \\
\hline
- & DENIS-P J025503.5--470050 & L8 & 4.9$\pm$0.3 & 1,2 \\
- & 2MASSW J1438082+640836 & L6 & 7.2 & 3 \\
- & 2MASSI J1507476--162738 & L5 & 7.33$\pm$0.03$^*$ & 2,4 \\
D2252 & DENIS-P J225210.7-173013.4 & L7.5 & 8.3$\pm$0.6 & 5 \\
- & 2MASSI J0835425--081923 & L5 & 8.3$\pm$0.9 & 2 \\
- & 2MASSW J2306292+154905 & L4 & 8.6 & 3 \\
- & 2MASSI J0036159+182110 & L3.5 & 8.76$\pm$0.06$^*$ & 3,6 \\
D1048 & DENIS-P J104842.8+011158.2 & L4 & 9.1$\pm$1.0 & 5 \\ 
- & 2MASSI J1515009+484739 & L6.5 & 9.2$\pm$1.8 & 2 \\
- & GJ 1001B (=LHS 102B) & L5 & 9.6$\pm$1.2$^*$ & 7,8 \\ 
- & 2MASSW J0045214+163445 & L3.5: & 10.4 & 9,10 \\
- & LSR0602+3910           & L1   & 10.6$\pm$0.8 & 9 \\
D1425 & DENIS-P J142528.0--365023.4 & L5 & 10.6$\pm$1.1 & 5 \\
D1705 & DENIS-P J170548.4--051645.7 & L4 & 10.7$\pm$1.0 & 5 \\
- & 2MASSW J0825196+211552 & L7.5 & 10.7 & 3 \\
- & 2MASSI J0439010--235308 & L6.5 & 10.8$\pm$1.1 & 2 \\
\hline
\end{tabular}
\end{center}
\begin{flushleft}
*  Distances from parallaxes in the given references. \\
1. Mart\'{\i}n et al. (1999) 2. \cite{cru03} \\
3. Bouy et al. (2003) 4. \cite{dah02} \\
5. This Letter 6. \cite{rei00} 7. \cite{v95} \\
8. \cite{e99} 9. \cite{sal03}\\
10. \cite{wil03} \\
\end{flushleft}
\end{table}

Spectral types have been estimated by direct comparison with known template objects. 
In Fig.~\ref{f1}, 
these results are shown over the whole wavelength range so employed. 
The ranges are constrained by the presence of telluric water vapour absorptions: 
we used $K$: 1.95--2.42\,$\mu$m and $H$: 1.45--1.8\,$\mu$m. The 
templates used have the following sources: \cite{geb02}: 2M\,1632 L8, SDSS\,2249 L5, 
SDSS\,0236 L6, Kelu-1 L2: \cite{rei01}: 
2M\,0036 L3.5, 2M\,0746 L0.5: \cite{leg01}: BRI\,0021 M9.5, LP\,944 M9, LHS\,429 
(=GL\,644C) M9, t513 (=TVLM\,513--46546) M8.5. All have been 
allocated spectral types on a common system (\cite{kir99}); hence our derived types are also on this 
system. While as many templates as possible were compared to the data, the grid of templates is not
uniform in terms of wavelength or spectral type coverage: we have plotted the best fits
to the data and adopted the spectral types so suggested (Fig.~\ref{f1}). In some cases it was found necessary to create 
a new template by averaging two template spectra, whose
types differ by no more than 1.5 subclasses, to yield a satisfactory fit to the data.
Where possible, independent types have been derived from $K$- and $H$-band spectra, in which cases the
spectral type quoted in Table 1 is a mean of the two estimates. Separate derivations from the $K$-
and $H$-bands are shown in Fig. ~\ref{f1}. 
 In almost all cases we find good agreement, to
$\pm$\,1 subclass, for objects which have both $K$- and $H$-band spectra, and a similar level of
agreement between our spectral types and those given for previously characterised
objects. Exceptions are D1048,
for which \cite{haw02} give L1 from far-red data, but for which our near-infrared data require 
a later type of L4, and the  
M9 object D1456, for which the $H$-band data suggest M7; in this case we adopt M9, as indicated by
the $K$-band data. The discrepancy in the former case might suggest binarity,
with the later-type component having a larger relative contribution to the $K$-band flux. 
A special case is the object 2MJ1112 typed in the $J$-band; for this star, the spectral type can only be
constrained to be later than M8.5 and earlier than L3.5. We adopt L0.5\,$\pm$\,2 subclasses. Distances
have been derived using the M$_{\rm J}$ vs. spectral type calibration of \cite{cru03}; errors on the
distance corresponding to $\pm$\,0.5\,subclasses in spectral type are given in Table 1. Of the known objects - 
apart from D1048 which may be rather later and closer than previously cited, distances have been derived using the previously published spectral types, with errors also representing $\pm$\,0.5\,subclasses.

\begin{table}
\begin{center}
\caption[] {Proper motions for the target objects derived from SuperCosmos and 2MASS positions 
with the epoch difference indicated. For typical positional errors $\sim$\,0.3\,\arcsec, uncertainties on the proper 
motions are therefore $\sim$\,100\,mas\,yr$^{-1}$, for
$\Delta_{\rm epoch}$\,=\,4\,yr.
Transverse velocities are derived using the distances
in Table 1 and the proper motion values, added in quadrature.}  
\begin{tabular}{lllllll}\hline\hline
Name & Sp.& $\mu_\alpha$cos$\delta$ & $\mu_\delta$ & $\Delta_{\rm epoch}$ & v$_{\rm trans}$ &  \\
     &    & mas\,yr$^{-1}$          &  mas\,yr$^{-1}$    & yr &           km\,s$^{-1}$ &           \\
\hline
D1048 & L4 &--210 & --190 & 4.916 & 12 \\ 
D1411$^a$ & M9 & --30 & +90    & 1.929  & 7 \\
D1425 & L5 & --260 & --470 & 8.063 & 27 \\
D1456 & M9 & --280 & --740 & 3.648 & 83 \\
D1510$^a$ & M8 & +80 & +50 & 1.877 & 9  \\
D1514$^b$ & M7 & - & - & - & - & \\
D1539 & L2 & +640 & +80 & 6.708 & 60  \\
D1705 & L4 & +100 & --130 & 6.741 & 8  \\
D2036 & M9.5 & +400 & --290 & 3.605 & 98  \\
D2057 & L1.5 & +60 & --210 & 3.211 & 17  \\
D2200 & L0 & +250 & --80 & 4.916 & 27 \\
D2229 & M9.5 & +170 & --30 & 4.251 & 32 \\
D2252 & L7.5 & +400 & +100 & 4.864 & 16  \\
D2254 & L0.5 & $\sim$\,0 & --30 & 5.909 & 3 \\
\hline
2MJ1112 & L0.5: & --450 & $\sim$\,0 & 7.066 & 88 & \\
\hline 
\end{tabular}
\end{center}
a. $\Delta_{\rm epoch}$ small b. Uncertain \\
\end{table}

In Table 2, we present our findings in the context of the currently known L-dwarf population in the 
immediate solar neighbourhood. It is clear that four mid-to-late L-type dwarfs presented here,
D1048, D1425, D1705, and D2252, have properties which strongly imply that they lie at distances of no more
than $\sim$\,10\,pc, if single. For D1048, our spectral type derivation of L4 brings this object newly into the $\la$\,10\,pc 
sample. 
Together, these objects represent a significant addition to the known sample of such dwarfs close 
to the Sun. We caution that, as is the case for many ultracool dwarfs, distances based on parallaxes are unknown
and the distances we determine rely on calibrations derived from larger samples of
field objects. We note also that the known object 2MASSI J0746425+200032 has a spectral type and photometry
which suggest a distance of only 9.5\,pc, which would place it in Table 2; yet, further study has revealed this 
object as a binary
(\cite{rei01}; \cite{bou03}), with d$_{\pi}$\,=\,12.2\,pc (Dahn et al. 2002). Moreover, we note that
2MASSI J0423485--041403, recently reclassified T0 and with $\pi$\,=\,65.9\,mas 
(G.R. Knapp et al., in preparation) also no longer belongs in the $\la$\,10\,pc sample.  
The nearby L-dwarfs
identified in this Letter are therefore excellent candidates to be inspected for binarity, 
or for giant planet companions. 

Additionally, in Table 3, we have calculated proper motions using positional differences in $\alpha$ and $\delta$
between 2MASS Final Release and SuperCosmos images taken at different epochs. $\Delta_{\rm epoch}$ represents the timeline
over which the proper motions have been derived. Where this quantity is small ($\la$\,2\,yr)
we do not consider the proper motions to be accurate. For one object, D1514, the
SuperCosmos position is uncertain; no stellar object exists in that database with a similar 
$I$-magnitude to that in the DENIS catalogue, within $\sim$\,10\,\arcsec. Lastly, we have derived 
transverse velocities using the spectroscopic distances given in Table 1 and the proper motion 
measurements. We find values broadly consistent with the kinematics of disk stars. However, two late-M objects,
D1456 and D2036, have rather high velocities. For D1456, a proper motion estimate of 
$\mu_\alpha$cos$\delta$\,=\,--340\,mas\,yr$^{-1}$ and $\mu_\delta$\,=\,--650\,mas\,yr$^{-1}$
is given in the SuperCosmos data, in agreement with our values to $\sim$\,20\%. Of the four very nearby objects,
D1425 has the largest proper motion, again in good agreement with SuperCosmos,
which gives $\mu_\alpha$cos$\delta$\,=\,--310\,mas\,yr$^{-1}$ and $\mu_\delta$\,=\,--440\,mas\,yr$^{-1}$. 
While the current data are not suitable for computing radial velocities, it is clear from our transverse
velocities that D1456 and D2036 are likely to belong to the dynamically old M-dwarf population, typified by
the template such object, Barnard's Star (GI\,699, M4V, v$_{\rm trans}$\,=\,89.5\,km\,s$^{-1}$). 2MJ1112
has a similarly high v$_{\rm trans}$, but we caution that its spectral type is not well enough constrained
to yet claim it as an L-dwarf member of this population.  

\section{Conclusions}

We present spectroscopic and kinematic data for 15 late M and  L-dwarfs, all but one taken from the DENIS catalogue. Spectral types have 
been determined by direct comparison to known L-type templates, in both $H$- and $K$-bands. Proper motions
derived by comparison of 2MASS and SuperCosmos positions yield transverse velocities consistent with membership of the
disk population, at least for all confirmed L-type objects. Three probable very late M-type objects have high transverse
velocities ($\sim$\,100\,km\,s$^{-1}$) and are likely to belong to a dynamically older thick disk population.
Nine objects in 
the sample are hitherto unpublished; three of these, and one further object, are shown to have have spectral 
types in the range L4--L7.5 and are relatively bright (17.9\,$<$\,$I$\,$<$\,16.2); 
hence, if single objects, they are extremely close, $\la$\,10\,pc. This last finding represents an increase in the
number of known L-dwarfs likely to be within $\sim$\,10\,pc of the Sun from 12 to 16.

\begin{acknowledgements}

The authors thank the observer, Scott Dahm, for preliminary data reduction.
Part of this work was carried out by TRK during a visit to the Institute for Astronomy, University of Hawaii,
funded by the National Science Foundation (NSF) grant AST-0205862. The publication makes use of data products 
from the Two Micron All Sky Survey, which is a joint project of the University of Massachusetts and the Infrared 
Processing and Analysis Centre/California Institute of Technology, funded by the National Aeronautics and Space 
Administration and the National Science Foundation. This research has made use of the SIMBAD database, operated at CDS,
Strasbourg, France. 
\end{acknowledgements}

\end{document}